\documentclass[a4paper,12pt]{article}
\usepackage{amsmath, amsthm, amsfonts, amssymb}

\numberwithin{equation}{section}

\begin{document}

\begin{center}
{\bf \large ON LOCAL PERTURBATIONS OF SHR\"ODINGER OPERATOR IN
AXIS}
\end{center}

\medskip
\begin{center}
Rustem R. GADYL'SHIN\footnote{The author is supported by RFBR
(grants No. 02-01-00768, 00-15-96038) and Ministry of Education of
RF (grant No. E00-1.0-53).}
\end{center}

\medskip

\begin{quote}
{ \emph{Bashkir State Pedagogical University, October Revolution
St.~3a, 450000, Ufa, Russia, E-mail:} \texttt{gadylshin@bspu.ru} }
\end{quote}

\medskip

\begin{center}
Abstract
\end{center}
\begin{quote}\quad
{\small We adduce the necessary and sufficient condition for
arising of eigenvalues of Shr\"odinger operator in axis under
small local perturbations. In the case of eigenvalues arising we
construct their asymptotics.}
\end{quote}

\bigskip

\centerline{\large\bf 1.~Introduction}
\medskip

The questions addresses the existence of bound states and the
asymptotics of associated eigenvalues (if they exist) for
Shr\"odinger operator with small potential in axis are have been
studied in [1]--[4]. The technique employed in these works based
on the self-adjointness of the perturbed equation. In present
paper it is considered a small perturbation which is arbitrary
localized second-order operator and the necessary and sufficient
conditions for arising of eigenvalues of perturbed operator are
adduced. In the case of eigenvalues arising we construct their
asymptotics. The main idea of the technique suggested giving a
simple explanation of "non-regular" (optional) arising of
eigenvalues under, obviously, regular perturbation  is as
follows. Instead of spectral parameter $\lambda$ we introduce
more natural frequency parameter $k$ related to spectral one by
the equality $\lambda=-k^2$, where $k$ lies in a complex
half-plane  $\mathrm{Re}\, k>0$. The solutions of both
non-perturbed and perturbed equations are extended w.r.t.
complex parameter on all complex plane. Under such extension the
solution of non-perturbed problem has a pole at zero that moves
under perturbation, while the residue at this pole (for both
non-perturbed and perturbed problems) is a solution of
corresponding homogeneous equation. For non-perturbed this
residue is a constant which is considered as exponent with index
$-kx$, where $k=0$. Depending on side to which this pole moves,
we obtain exponential increasing or decreasing residue for
perturbed problem. As a result, if pole moves into the
half-plane $\mathrm{Re}\, k>0$ then the eigenvalue arises, while
pole moving to the half-lane $\mathrm{Re}\, k\le0$ do not
produce pole. The direction of moving is determined by the
operator of perturbation.

The structure of the paper is as follows. In the second section
we state the main result, in the third is adduced its proof. In
the fourth section we demonstrate some examples illustrating the
main statement of the paper.

\bigskip

\begin{center} {\large\bf 2.~Formulation of the main result}
\end{center}

\medskip

Hereinafter $W_{2,loc}^j({\mathbb{R}})$ is a set of functions
defined on ${\mathbb{R}}$ whose restriction to any bounded
domain $D\subset {\mathbb{R}}$ belongs to $W_2^j(D)$,
$\|\bullet\|_G$ and $\|\bullet\|_{j,G}$ are norms in $L_2(G)$
and $W_2^j(G)$, respectively. Next, let $Q$ be an arbitrary
fixed interval in ${\mathbb{R}}$, $L_2({\mathbb{R}};Q)$ be the
subset of functions in $L_2({\mathbb{R}})$ with supports in
$\overline {Q}$, ${\mathcal L}_\varepsilon$ be linear operators
mapping  $W_{2,loc}^j({\mathbb{R}})$ into $L_2({\mathbb{R}};Q)$
such that $\|{\mathcal L}_\varepsilon[u]\|_{Q}\le C({\mathcal
L})\,\|u\|_{2,Q}$, where constant $C({\mathcal L})$ does not
depends on $\varepsilon$, $0<\varepsilon<<1$,
$$
\begin{array}{l}
\left<g\right>=\int\limits_{-\infty}^\infty g\,dx,\qquad
H_0=-\frac{d^2}{d x^2},\qquad H_\varepsilon=-\left(\frac{d^2}{d
x^2}+\varepsilon{\mathcal L}_\varepsilon\right).
\end{array}
$$
We define linear operators $A(k)\,:\,L_2({\mathbb{R}};Q)\to
W^2_{2,loc}({\mathbb{R}})$ and
$T_\varepsilon^{(0)}(k)\,:\,L_2({\mathbb{R}};Q)\to
L_2({\mathbb{R}};Q)$ in the following way:
$$
A(k)g=- \frac{1}{2k}\int\limits_{-\infty}^\infty e^{-k|x-t|}
g(t)\,dt,\qquad T_\varepsilon^{(0)}(k)g={\mathcal
L}_\varepsilon[A(k)g]+\frac{\left<g\right>}{2k} {\mathcal
L}_\varepsilon[1].
$$
Denote by $\mathcal{B}(X,Y)$ the Banach space of linear bounded
operators mapping Banach space $X$ into Banach space $Y$,
$\mathcal {B}(X)\overset{def}{=}\mathcal{B}(X,Y)$. We indicate by
$\mathcal{B}^h(X,Y)$ (by $\mathcal{B}^h(X)$)  the set of
holomorphic operator-valued functions whose values belongs to
$\mathcal{B}(X,Y)$ (to $\mathcal{B}(X)$). We use the notation $I$
for identity mapping and  the notation $S^t$ for a circle in
$\mathbb{C}$ of radius $t$ with center at zero. Since by
definition of $T_\varepsilon^{(0)}(k)$ we have that
$T_\varepsilon^{(0)}(k)\in \mathcal{B}^h(L_2(\mathbb{R};Q))$,
$$
T_\varepsilon^{(0)}(k)g= \frac{1}{2}{\mathcal
L}_\varepsilon\left[\int\limits_{-\infty}^\infty
g(t)|x-t|\,dt\right]+kT_\varepsilon^{(1)}(k)g,\qquad
T_\varepsilon^{(1)}(k)\in \mathcal{B}^h(L_2(\mathbb{R};Q)),
$$
then we arrive at the following statement.

{\bf Lemma 2.1.} {\it Let $S_\varepsilon(k)=(I+\varepsilon
T_\varepsilon^{(0)}(k))^{-1}$. Then for all $R>0$ there exist
$\varepsilon_0(R)>0$, such that for
$\varepsilon<\varepsilon_0(R)$ and $k\in S^R$ the
operator-valued function $S_\varepsilon(k)\in
\mathcal{B}^h(L_2(\mathbb{R};Q))$,
$S_\varepsilon(k)\underset{\varepsilon\to0}{\to}I$ uniformly on
$k$, and the equation
$$
k-\frac{\varepsilon}{2}\left<S_\varepsilon(k){\mathcal
L}_\varepsilon[1]\right> =0\eqno(2.1)
$$
has a unique solution $k_\varepsilon\in S^R$, and also,
$$
k_\varepsilon=\varepsilon\frac{1}{2}\left(m_\varepsilon^{(1)}+\varepsilon
m_\varepsilon^{(2)}+O(\varepsilon^2)\right),\eqno(2.2)
$$
where
$$m_\varepsilon^{(1)}=\left<{\mathcal
L}_\varepsilon[1]\right>,\qquad
m_\varepsilon^{(2)}=-\int\limits_{-\infty}^\infty{\mathcal
L}_\varepsilon \left[\int\limits_{-\infty}^\infty|x-y|{\mathcal
L}_\varepsilon[1](y)\,dy\right](x) \,dx.\eqno(2.3)
$$
}

Let us call the operator $\mathcal{L}_\varepsilon$ the real one,
if $\mathrm{Im}\,<\overline{g}\mathcal{L}_\varepsilon[g]>=0$ for
all $g\in W_{2,loc}^2(\mathbb{R})$. We denote
$\Pi_s(t)=\{k:\,|\mathrm{Im}\,\,k|< sC({\mathcal
L}),\,\,\mathrm{Re}\,k>t\}$, and we indicate by
$\Sigma(H_\varepsilon)$ the set of eigenvalues of operator
$H_\varepsilon$. The aim of this paper is to prove the following
statement.

{\bf Theorem 2.1.} {\it If $\mathrm{Re}\, k_\varepsilon\le0$,
then there exist
$t(\varepsilon)\underset{\varepsilon\to0}{\to}\infty$, such that
$\Sigma(H_\varepsilon)\subset \Pi_\varepsilon(t(\varepsilon))$.
If, in addition, the operator  ${\mathcal L}_\varepsilon$ is
real, then $\Sigma(H_\varepsilon)\subset
(t(\varepsilon),\infty)$.

If $\mathrm{Re}\, k_\varepsilon>0$, then there exist
$t(\varepsilon)\underset{\varepsilon\to0}{\to}\infty$, such that
$\Sigma(H_\varepsilon)\backslash
\Pi_\varepsilon(t(\varepsilon))=\{\lambda_\varepsilon\}$,
$$
\lambda_\varepsilon=-k_\varepsilon^2,\eqno(2.4)
$$
and the associated single eigenfunction $\phi_\varepsilon$ has
the form
$$
\phi_\varepsilon=A(k_\varepsilon)S_\varepsilon(k_\varepsilon){\mathcal
L}_\varepsilon[1].\eqno(2.5)
$$
If, in addition, the operator ${\mathcal L}_\varepsilon$ is
real, then $\Sigma(H_\varepsilon)\backslash
(t(\varepsilon),\infty)=\{\lambda_\varepsilon\}$.
}

{\bf Remark 2.1.} The statements of Theorem 2.1 does not
excludes the situation when $\mathrm{Re}\,k_\varepsilon\le0$ for
some values of $\varepsilon$ and $\mathrm{Re}\,k_\varepsilon>0$
for other those of $k_\varepsilon$ (see example 4.3).

Directly from Lemma 2.1 (namely, from equation (2.1)) and
Theorem 2.1 it follows

{\bf Corollary 2.1.} {\it If ${\mathcal
L}_\varepsilon[1]\equiv0$, then there exists
$t(\varepsilon)\underset{\varepsilon\to0}{\to}\infty$, such that
$\Sigma(H_\varepsilon)\subset \Pi_\varepsilon(t(\varepsilon))$.
If, in addition, the operator ${\mathcal L}_\varepsilon$ is
real, then $\Sigma(H_\varepsilon)\subset
(t(\varepsilon),\infty)$.}

\bigskip

\begin{center}
{\large\bf 3.~Proof of Theorem 2.1}
\end{center}

\medskip

Let us denote by $\mathcal{B}^m(X,Y)$ (by $\mathcal{B}^m(X)$)
the set of meromorphic operator-valued functions with values in
$\mathcal{B}(X,Y)$ (â $\mathcal{B}(X)$). The set of linear
operators mapping Banach space $X$ into
$W_{2,loc}^2(\mathbb{R})$ such that their restriction to any
bounded set $D$ belongs to $\mathcal{B}(X,W_{2}^2(D))$ is
indicated by $\mathcal{B}(X,W_{2,loc}^2(\mathbb{R}))$.
Similarly, we use the notation
$\mathcal{B}^h(X,W_{2,loc}^2(\mathbb{R}))$ (
$\mathcal{B}^m(X,W_{2,loc}^2(\mathbb{R}))$) for the set of
operator-valued functions with values in
$\mathcal{B}(X,W_{2,loc}^2(\mathbb{R}))$ such that for all
bounded  $D$ they belongs to $\mathcal{B}^h(X,W_{2}^2(D))$ (to
$\mathcal{B}^m(X,W_{2}^2(D))$). Next, let $P_\varepsilon(k)$ be
the operator defined by the equality
$$
P_\varepsilon(k)f=\varepsilon\frac{\left<S_\varepsilon(k)f\right>
S_\varepsilon(k)\mathcal{L}_\varepsilon[1]}
{2k-\varepsilon\left<S_\varepsilon(k)\mathcal{L}_\varepsilon[1]\right>}+
S_\varepsilon(k)f,
$$
$\mathcal{R}_\varepsilon(k)\overset{def}{=}A(k)P_\varepsilon(k)$,
$ \mathbb{C}_+\overset{def}{=}\{z:\,\mathrm{Re}\, z>0\}$.

{\bf Theorem 3.1.} {\it For all $R>0$ there exists
$\varepsilon_0(k)>0$ such that
\begin{enumerate}
\def\theenumi{\arabic{enumi})}
\item
$\mathcal{R}_\varepsilon(k)\in
\mathcal{B}^m(L_2(\mathbb{R};Q),W_{2,loc}^2(\mathbb{R}))$ as
$\varepsilon<\varepsilon_0$ and $k\in S^R$, and also, in $S^R$
there is the only pole $k_\varepsilon$ being a solution of the
equation (2.1) and it is a first order pole; if, in addition,
$k\in \mathbb{C}_+$, then $\mathcal{R}_\varepsilon(k)\in
\mathcal{B}^m(L_2(\mathbb{R},Q);W_{2}^2(\mathbb{R}))$;

\item for all $f\in L_2(\mathbb{R};Q)$ the function
$u_\varepsilon=\mathcal{R}_\varepsilon(k)f$ is a solution of the
equation
$$
-H_\varepsilon u_\varepsilon= k^2u_\varepsilon+f\quad\hbox{â
${\mathbb{R}}$};\eqno(3.1)
$$

\item the residue of the function $u_\varepsilon$ at the pole
$k_\varepsilon$ is defined by the equality (2.5) up to a
multiplicative factor, moreover, this factor is nonzero if
$\left<f\right>\not=0$.
\end{enumerate}
 }

{\bf Proof.} By definition,  $A(k)\in
\mathcal{B}^m(L_2(\mathbb{R};Q),W_{2,loc}^2(\mathbb{R}))$, and
also, $A(k)$ has a unique pole of first order at zero and
$A(k)\in \mathcal{B}^h(L_2(\mathbb{R};Q),W_{2}^2(\mathbb{R}))$
for $k\in \mathbb{C}_+$. Then bearing in mind the definition of
$\mathcal{R}_\varepsilon(k)$ and Lemma 2.1, we get consecutively
$\mathcal{R}_\varepsilon(k)$ having no pole at zero and validity
of statement 1) of Theorem being proved.

Let us proceed to the proof of the statement 2). We seek the
solution of the equation (3.1) in the form
$$
u_\varepsilon=A(k)g_\varepsilon,\eqno(3.2)
$$
where $g_\varepsilon$ is some function belonging to
$L_2({\mathbb{R}};Q)$. Substituting (3.2) into (3.1), we deduce
that  (3.2) is a solution of  (3.1) in the case
$$
(I+\varepsilon T_\varepsilon(k))g_\varepsilon=f,\eqno(3.3)
$$
where
$$
T_\varepsilon(k)={\mathcal L}_\varepsilon A(k).\eqno(3.4)
$$
It follows from (3.4) and the definition of ${\mathcal
L}_\varepsilon$ and $A(k)$ that the result of the action of the
operator $T_\varepsilon(k)$ is as follows:
$$
T_\varepsilon(k)g=-\frac{\left<g\right>}{2k} {\mathcal
L}_\varepsilon[1]+T_\varepsilon^{(0)}(k)g. \eqno(3.5)
$$

Let $R>0$ be an arbitrary number and $\varepsilon$ satisfies all
assumptions of Lemma 2.1. Applying the operator
$S_\varepsilon(k)$ to both hands of the equation (3.3) and
taking into account (3.5), we obtain that
$$
\left(g_\varepsilon-\varepsilon\frac{\left<g_\varepsilon\right>}{2k}S_\varepsilon(k){\mathcal
L}_\varepsilon[1]\right)= S_\varepsilon(k)f.\eqno(3.6)
$$
Having integrated (3.6), we deduce
$$
\left<g_\varepsilon\right>\left(1-\frac{\varepsilon}{2k}\left<S_\varepsilon(k){\mathcal
L}_\varepsilon[1]\right>
\right)=\left<S_\varepsilon(k)f\right>.\eqno(3.7)
$$
The equality (3.7) allows us to determine
$\left<g_\varepsilon\right>$; substituting its value into (3.6),
we easily get the formula
$$
g_\varepsilon=P_\varepsilon(k)f.\eqno(3.8)
$$
The assertions  (3.2) and (3.8) yield the validity of the
statement 2). In its turn, the correctness of  statement 3) is
the implication from  1) and 2) and the definition of $\mathcal
{R}_\varepsilon(k)$. The proof is complete.

We will use the notation $R_\varepsilon(\lambda)$ for the
resolvent of the operator $H_\varepsilon$. It is well known fact
that the set of eigenvalues coincide with the set of poles of
the resolvent, while the coefficient of the pole (of highest
order) is a projector into the space that is a span of
eigenfunctions associated with this eigenvalue.

{\bf Lemma 3.1.} {\it The number of poles of the resolvent
$R_\varepsilon(\lambda)$, their orders and the dimensions of the
residues at them are completely determined by the functions
belonging to $L_2(\mathbb{R};Q)$}.

{\bf Proof.} Let $F$ be an arbitrary function with compact support
$D$. There is no loss of generality in assuming that $\{0\}\in Q$.
We use symbols  $R_+(\lambda)$ and $R_-(\lambda)$ for the
resolvents of the Dirichlet problem for $H_0$ in the positive
$\mathbb{R}^+$ and negative $\mathbb{R}^-$ real semi-axises
respectively, by $F_+$ and $F_-$ we denote the restrictions of $F$
to these axises. We use symbol $\chi\in C^\infty(\mathbb{R})$ for
the cut-off function vanishing in a neighbourhood of zero and
equalling to one outside $Q$. Let $\mathbb{R}_+$ be the
nonnegative imaginary semi-axis, we also set
$\mathbb{C^+}=\mathbb{C_+}\cup \mathbb{R}_+$. Since the function
$\lambda=-k^2$ establishes one-to-one correspondence from
$\mathbb{C}^+$ onto $\mathbb{C}$, then for  $\lambda\in
\mathbb{C}$ (or, equivalently, for $k\in \mathbb{C^+}$)
$$
R_\pm(\lambda)F_\pm(x)=R_\pm(-k^2)F_\pm(x)=
\pm\frac{1}{2k}\int\limits_{0}^{\pm\infty}
\left(e^{-k|x-t|}-e^{-k|x+t|}\right) F(t)\,dt.
$$
On the other hand,
$$
U_\pm(x;k)=\pm\frac{1}{2k}\int\limits_{0}^{\pm\infty}
\left(e^{-k|x-t|}-e^{-k|x+t|}\right) F(t)\,dt
$$
are holomorphic functions in $\mathbb{C}$ with values in
$W_{2,loc}^2(\mathbb{R}^\pm)$ (i.e., their restrictions to all
bounded domains $G$ are holomorphic functions with values
belonging to $W_{2,loc}^2(G)$). For this reason the function
$\chi (R_+(-k^2)F_++R_-(-k^2)F_-)$ can be extended in
$\mathbb{C}$ as holomorphic function with values in
$W_{2,loc}^2(\mathbb{R})$. The solution of the equation
$$
(H_\varepsilon-\lambda)U=F\quad \hbox{â $\mathbb{R}$}\eqno(3.9)
$$
is sought in the form
$$
U=u+\chi (R_+(-k^2)F_++R_-(-k^2)F_-),\eqno(3.10)
$$
where $-k^2=\lambda$, $k\in \mathbb{C^+}$. Substituting  (3.10)
into (3.9), we obtain the equation $(H_\varepsilon-\lambda)u=f$
for $u$, where the function $f(x;\lambda)=f(x;-k^2)$ can be
extended w.r.t. $k$ in $\mathbb{C}$, that is a holomorphic
function with values in $L_2(\mathbb{R};Q)$. Since the second
term in (3.10) can be extended holomorphically in $\mathbb{C}$,
then it implies the validity of the lemma being proved.

{\bf Theorem 3.2.} {\it Let $R>0$ be an arbitrary number,
$\varepsilon_0$ and $k_\varepsilon$ to satisfy Theorem 3.1,
$\lambda=-k^2$. Then
$$
\mathcal{R}_\varepsilon(k)f=-R_\varepsilon(\lambda)f\eqno(3.11)
$$
for all $f\in L_2(\mathbb{R};Q)$ and for all $k\in
\mathbb{C}^+\cap S^R$ (or, equivalently, for all $\lambda\in
S^{R^2}$).

If $\mathrm{Re}\, k_\varepsilon\le0$, then
$\Sigma(H_\varepsilon)\cap S^{R^2}=\emptyset$.

If $\mathrm{Re}\, k_\varepsilon>0$, then
$\Sigma(H_\varepsilon)\cap S^{R^2}=\{\lambda_\varepsilon\}$,
where $\lambda_\varepsilon$ and the associated single
eigenfunction are determined by the equalities (2.4) and (2.5).}

\textbf{Proof.}  Since the function $\lambda=-k^2$ establishes
one-to-one correspondence from $\mathbb{C}^+\cap S^R$ onto
$S^{R^2}$, then the validity of the equality (3.11) follows from
the statement 2) of Theorem 3.1 and the definition of the
resolvent. The correctness of the rest statement of the theorem
begin proved follows from Theorem 3.1 and Lemma 3.1. The proof is
complete.

{\bf Lemma 3.2.} {\it
$\Sigma(H_\varepsilon)\subset\Pi_\varepsilon(-\varepsilon
C({\mathcal L}))$. If the operator ${\mathcal L}_\varepsilon$ is
real, then $\Sigma(H_\varepsilon)\subset [-\varepsilon
C({\mathcal L}),\infty)$.}

{\bf Proof.} Let
$$
\lambda_\varepsilon\in\Sigma(H_\varepsilon)\backslash\left(
\mathbb{R}^+\cup\{0\}\right).\eqno(3.12)
$$
Since $H_\varepsilon u=H_0u$ outside $\overline {Q}$, then there
exists normalized in $L_2(\mathbb{R})$ function
$\phi_\varepsilon\in W_2^2(\mathbb{R})$, such that
$$
H_\varepsilon\phi_\varepsilon=\lambda_\varepsilon
\phi_\varepsilon.\eqno(3.13)
$$
Multiplying both hands of (3.13) by $\overline{\phi_\varepsilon}$
and integrating by part, we obtain the equality
$$
\|\phi'_\varepsilon\|^2_{\mathbb{R}}-\varepsilon\left<\overline{\phi_\varepsilon}
{\mathcal L}_\varepsilon\phi_\varepsilon\right>=
\lambda_\varepsilon.\eqno(3.14)
$$
Calculating the real and imaginary part of (3.14), employing the
estimate $\|{\mathcal L}_\varepsilon\phi_\varepsilon\|_Q\le
C({\mathcal L})\|\phi_\varepsilon\|_{2,Q}$ and bearing in mind
(3.12), we conclude the statement of the lemma being proved is
true.

It is easily seen that Theorem  2.1 is a direct implication of
Theorem 3.2 and Lemma 3.2.

\bigskip

\centerline{\large\bf 4.~Examples}

\medskip

{\bf Example 4.1.} Let $\mathcal{L}_\varepsilon[g]=Vg$, where
$V\in C^\infty_0(Q)$. Then in view of  (2.2), (2.3) and Theorem
2.1 we obtain, that an inequality
$\mathrm{Re}\,\left<V\right><0$ yields
$\mathrm{Re}\,k_\varepsilon<0$ and, therefore, the operator
$H_\varepsilon$ has no eigenvalue converging to zero, while
opposite inequality $\mathrm{Re}\,\left<V\right>\,>0$ implies
that such eigenvalue exists and satisfies the asymptotics
$$
\lambda_\varepsilon=-\varepsilon^2\frac{\left<V\right>}{4}+O(\varepsilon^3).\eqno(4.1)
$$

In the case when $\left<V\right>=0$, taking into account that
(in this case)
$$
-\int\limits_{-\infty}^\infty
\int\limits_{-\infty}^\infty|x-y|V(x)V(y)\,dydx=
2\int\limits_{-\infty}^\infty\left(\int\limits_{-\infty}^xV(y)dy\right)^2
dx,\eqno(4.2)
$$
by the assertion (2.3), we get that
$$
m_\varepsilon^{(1)}=<V>=0,\qquad
m_\varepsilon^{(2)}=2\int\limits_{-\infty}^\infty
\left(\int\limits_{-\infty}^xV(y)dy\right)^2dx.\eqno(4.3)
$$
If, in addition, $\mathrm{Im}\,V=0$, then due to (4.1), (4.3) and
Theorem 2.1 the eigenvalue exists and has the asymptotics
$$
\lambda_\varepsilon=-\varepsilon^4\left(\int\limits_{-\infty}^\infty
\left(\int\limits_{-\infty}^xV(y)dy\right)^2dx\right)^2+O(\varepsilon^5).
\eqno(4.4)
$$
For real $V$ the asymptotics  (4.1), (4.4) have been derived in
[1]. So, the asymptotics  (4.1) is a generalization for the case
of complex-valued potentials. Observe that for real  $V$ the
inequality $\left<V\right>\ge0$ is an necessary and sufficient
condition of the existence of eigenvalue of $H_\varepsilon$
(what was proved in [1]). However, if $V$ is a complex-valued
function, then the assumption $\left<V\right>=0$ is not
sufficient for the existence of the eigenvalue. Indeed, it is
easy to see that if  $V=u'+i2u'$, where $u\in C^\infty_0(Q)$ is
a real function then
$$
m_\varepsilon^{(1)}=<V>=0,\qquad
\mathrm{Re}\,m_\varepsilon^{(2)}=-6\int\limits_{-\infty}^\infty
u^2(x)dx<0,
$$
and by the assertion (2.2) and Theorem 2.1 the eigenvalue of
$H_\varepsilon$ does not exist.

{\bf Example 4.2.} Let $\mathcal{L}_\varepsilon[g]=V_\varepsilon
g$, where $V_\varepsilon=V+\varepsilon V_1$, and  $V,\,V_1$ are
real functions with supports in $Q$. Due to (2.2), (2.3), (4.2)
and Theorem 2.1 the condition $\left<V_\varepsilon\right>\ge0$
is sufficient for existence of eigenvalue which has the
asymptotics (4.1) as $\left<V\right>\,>0$ and the asymptotics
$$
\lambda_\varepsilon=-\varepsilon^4\left(\frac{1}{2}\left<V_1\right>+
\int\limits_{-\infty}^\infty
\left(\int\limits_{-\infty}^xV(y)dy\right)^2dx\right)^2+O(\varepsilon^5),
$$
as $\left<V\right>=0$. However, in distinction to classic case
(real $V_\varepsilon=V$) the condition
$\left<V_\varepsilon\right>\,<0$ is not sufficient for absence
of eigenvalue. Indeed if $Q=(-\pi/2,\pi/2)$ and
$V_\varepsilon=\sin x-\varepsilon\cos x$ then in view of  (2.2),
(2.3) $\left<V_\varepsilon\right>=-2<0$, but
$k_\varepsilon=\varepsilon^2\frac{\pi-2}{2}+O(\varepsilon^3)$.
Hence, by Theorem 2.1 the eigenvalue exists.

{\bf Example 4.3.} Let
$\mathcal{L}_\varepsilon[g]=\exp\{i\varepsilon^{-1}\}Vg$, where
$V\in C^\infty_0(Q)$ is a real function and $\left<V\right>\,>0$.
Then by (2.2), (2.3) and Theorem 2.1 we obtain that, for all
sufficient large $n$ and any $0<\delta<\pi/2$, the eigenvalues are
absent as $\left(3\pi/2+2\pi n-\delta\right)^{-1}
<\varepsilon<\left(\pi/2+2\pi n+\delta\right)^{-1}$ while an eigenvalue
exists as $\left(\pi/2+2\pi
n-\delta\right)^{-1}<\varepsilon<\left(-\pi/2+2\pi
n+\delta\right)^{-1}$ for each fixed $\delta>0$ and satisfies the
asymptotics
$$
\lambda_\varepsilon=-\left(\varepsilon\cos\frac{1}{\varepsilon}\right)^2\frac{1}{4}
\left<V\right>^2+O(\varepsilon^3).
$$

{\bf Example 4.4.} Let
$$
\mathcal{L}_\varepsilon=a_2\frac{d^2}{dx^2}+a_1\frac{d}{dx}+
V_\varepsilon,
$$
where $a_j,\,V_\varepsilon\in C^\infty_0(Q)$. Since
$\mathcal{L}_\varepsilon[1]=V_\varepsilon$, then in the case
$V_\varepsilon\equiv0$ an eigenvalue is absent due to Corollary
2.1 and the equality
$\left<\mathcal{L}_\varepsilon[1]\right>=\left<V_\varepsilon\right>$
implies that

1) if $V_\varepsilon=V$ and $\mathrm{Re}\, V\not=0$, then
$k_\varepsilon$ has asymptotics derived in Example 4.1;

2) if $V_\varepsilon=\exp\{i\varepsilon^{-1}\}V$ and
$\left<V\right>\,>0$, then $k_\varepsilon$ has asymptotics derived
in Example 4.3.

{\bf Example 4.5.} Let
$\mathcal{L}_\varepsilon[g]=\varkappa(Q)\left<\rho g\right>
$
where $\rho\in C^\infty_0(Q)$, and $\varkappa(Q)$ is a
characteristic function for $Q$, (i.e., this function equals to
one for $x\in Q$ and vanishes for other $x$). Then by (2.2),
(2.3), Theorem 2.1 ad Corrolary 2.1 an eigenvalue is absent if
$\left<\rho\right>=0$ or $\mathrm{Re}\,\left<\rho\right><0$,
and, if $\mathrm{Re}\,\left<\rho\right>\,>0$, then an eigenvalue
exists and has asymptotics
$$
\lambda_\varepsilon=-\varepsilon^2\frac{1}{4}\left(|Q|\left<\rho\right>\right)^2
+O(\varepsilon^3).\eqno(4.5)
$$

{\bf Example 4.6.} Let
$$\mathcal{L}_\varepsilon[g]=\varkappa(Q)\int\limits_{-\infty}^x
\rho(t) g(t)\,dt,
$$
where $\rho\in C^\infty_0(Q)$. Then the assertions (2.2), (2.3)
imply
$$
k_\varepsilon=\varepsilon\frac{1}{2}\left(|Q|\left<\rho\right>-
\left<x\rho\right>\right) +O(\varepsilon^2),\eqno(4.6)
$$
and, therefore, the eigenvalue exists if
$\left(|Q|\left<\rho\right>-\left<x\rho\right>\right)>0$, and it
is absent if
$\left(|Q|\left<\rho\right>-\left<x\rho\right>\right)<0$. In
particular, if $\rho$ is an even function and
$\left<\rho\right>\,>0$, then due to (2.4), (4.6) the eigenvalue
has asymptotics  (4.5).

\bigskip

\centerline{\large\bf References}

\medskip

\begin{enumerate}
\def\theenumi{[\arabic{enumi}]}

\item {\it B.~Simon.} Ann. Phys. 1976. V.~97. P.~279.

\item {\it M.~Klaus.} Ann. Phys. 1977. V.~108. P.~288.

\item {\it R.~Blankenbecler, M.L.~Goldberger,
B.~Simon.} Ann. Phys. 1977. V.~108. P.~69.

\item {\it M.~Klaus, B.~Simon.} Ann. Phys. 1980. V.~130.
P.~251.

\end{enumerate}
\end{document}